\documentclass[letter]{aa} 

\usepackage{color}
\usepackage{multirow}
\usepackage{mathrsfs}

\usepackage{graphicx} 
\usepackage{txfonts}
\def\simgr{\,\hbox{\hbox{$ > $}\kern -0.8em \lower 1.0ex\hbox{$\sim$}}\,}
\def\simle{\,\hbox{\hbox{$ < $}\kern -0.8em \lower 1.0ex\hbox{$\sim$}}\,}
\def\cpd{CPD\,$-$62$^{\circ}$\,2124}
\def\bzv{$\langle$B$_z\rangle$}
\def\bzn{$\langle$B$_z\rangle$}

\begin{document}

   \title{B field in OB stars (BOB): The outstandingly strong magnetic field in the evolved He-strong star \cpd\thanks{Based on observations made with ESO telescopes at the La Silla and Paranal observatories under programme ID 191.D-0255(G,I).}}
   \author{N.~Castro\inst{1,2}, 
           L.~Fossati\inst{3,2},
           S.~Hubrig\inst{4},
           S.~P.~J\"arvinen\inst{4},
           N.~Przybilla\inst{5},
           M.-F.~Nieva\inst{5},
           I.~Ilyin\inst{4},
           T.~A.~Carroll\inst{4},
           M.~Sch\"oller\inst{6},
           N.~Langer\inst{2},
           F.~R.~N.~Schneider\inst{7},
           S.~Simón-Díaz\inst{8,9},
           T.~Morel\inst{10},
           K.~Butler\inst{11}
            and the BOB collaboration. }
         \institute{Department of Astronomy, University of Michigan, 1085 S. University Avenue, Ann Arbor, MI 48109-1107, USA
              \email{ncastror@umich.edu}
         \and
         Argelander-Institut f\"ur Astronomie der Universit\"at Bonn, Auf dem H\"ugel 71, 53121, Bonn, Germany
         \and
         Space Research Institute, Austrian Academy of Sciences, Schmiedlstrasse 6, A-8042 Graz, Austria
         \and           
         Leibniz-Institut für A        strophysik Potsdam (AIP), An der Sternwarte 16, D-14482 Potsdam, Germany
                 \and 
         Institut f\"ur Astro- und Teilchenphysik, Universit\"at Innsbruck, Technikerstr. 25/8, 6020 Innsbruck, Austria 
         \and 
         European Southern Observatory, Karl-Schwarzschild-Str.~2, 85748 Garching bei M\"unchen, Germany 
         \and 
          Department of Physics, University of Oxford, Denys Wilkinson Building, Keble Road, Oxford OX1 3RH, United Kingdom      
          \and         
         Instituto de Astrofísica de Canarias, 38200, La Laguna, Tenerife, Spain 
         \and       
         Universidad de La Laguna, 38205, La Laguna, Tenerife, Spain 
          \and          
         Space sciences, Technologies and Astrophysics Research (STAR) Institute, Universit\'e de Li\`ege, Quartier Agora, All\'ee du 6 Ao\^ut 19c, B\^at. B5C, B4000-Li\`ege, Belgium
        \and 
        Universit\"ats-Sternwarte M\"unchen, Scheinerstr. 1, 81679 M\"unchen, Germany}
   \date{Received --; accepted --}
\titlerunning{The strong magnetic star \cpd}
\authorrunning{N. Castro et al.}
 
  \abstract
   {  The origin and evolution of magnetism in OB stars is far from being well understood. With approximately 70 magnetic OB stars known, any new object with unusual characteristics may turn out to be a key piece of the puzzle. We report the detection of an exceptionally strong magnetic field in the He-strong B2IV star CPD\,$-62^{\circ}2124$. Spectropolarimetric FORS2 and HARPSpol observations were analysed by two independent teams and procedures,  concluding on a strong longitudinal magnetic field of approximately $5.2$\,kG. The quantitative characterisation of the stellar atmosphere yields an effective temperature of 23650$\pm$250\,K, a surface gravity of 3.95$\pm$0.10\,dex and a surface helium fraction of 0.35$\pm$0.02 by number. The metal composition is in agreement with the cosmic abundance standard, except for Mg, Si and S, which are slightly non-solar. The strong and broad ($\sim 300\,$km\,s$^{-1}$) disc-like emission displayed by the H$\alpha$ line suggests a centrifugal magnetosphere supported by the strong magnetic field. Our results imply that CPD\,$-62^{\circ}2124$ is an early B-type star hosting one of the strongest magnetic fields discovered to date, and one of the most evolved He-strong stars known, with a fractional main-sequence lifetime of approximately 0.6.}

   \maketitle

\section{Introduction}

The recent efforts invested in the search for magnetic fields in the most massive  O- and B-type stars have provided more than 70 confirmed magnetic detections \citep[e.g.][]{2013MNRAS.429..398P,2014arXiv1404.5508A,2014A&A...562A.143F,2014A&A...564L..10H,2014A&A...563L...7N,2015arXiv150703591C} and a magnetic incidence rate of approximately 7\% \citep[][]{2013arXiv1310.3965W,2015A&A...582A..45F,2016arXiv161007895G,2016arXiv161104502S}. Any new addition to the scarce number of known magnetic stars, especially objects with unusual characteristics, has the potential to improve our understanding of the origin of magnetism and its role in stellar structure and evolution \citep{2012ARA&A..50..107L,2015A&A...584A..54P}, as well as in the characteristics of the circumstellar environment \citep{2013MNRAS.429..398P} and   spectral features linked to the magnetic field \citep[e.g. Of?p stars;][]{2010A&A...520A..59N}. Detections of strong (i.e. longitudinal field larger than 1 kG) magnetic fields have been reported in the literature among  He-strong stars \citep[e.g.][]{1978ApJ...224L...5L,1979ApJ...228..809B,1988PhDT........53B,2006A&A...450..777B,2015A&A...578L...3H},  pointing out He-strong stars as a  sub-class of strong magnetic early B-type stars.

Within the context of the ``B fields in OB stars'' (BOB) collaboration \citep{2014arXiv1408.2100M,2015IAUS..307..342M,2014A&A...564L..10H,2015A&A...578L...3H,2015A&A...582A..45F,2016arXiv161104502S}, here we report  the detection of an exceptionally strong magnetic field in the He-strong B2IV star \cpd\ \citep[$V$=11.04\,mag;][]{1981ApJ...250..701D,1983ApJ...268..195W,1997A&A...324..949Z,2009A&A...498..961R}. The star was observed  using two different instruments (Sect. \ref{SECT:osb}) and the magnetic field was detected and confirmed by independent teams employing different techniques (Sect. \ref{SECT:Mag}). Section \ref{SECT:Para} presents the stellar atmospheric and chemical abundance analyses, and gives the derived stellar evolution properties. The results are discussed and conclusions are drawn  in Sect. \ref{Ha}.

\section{Observations}
\label{SECT:osb}

We observed \cpd\ on the 17th March, 2015, using the FORS2 low-resolution spectropolarimeter \citep{1998Msngr..94....1A} attached to the ESO/VLT UT1 (Antu) of the Paranal Observatory (Chile). The data were taken with a slit width of 0.4" and the grism 600B. This setting led to a resolving power of approximately 1700 and a wavelength coverage from 3250-6215\,\AA. Eight consecutive exposures of 600 seconds were carried out with a total exposure time of 4800 seconds. More details on the instrument settings and exposure sequence are provided by \cite{2014A&A...564L..10H} \citep[see also][]{2015A&A...582A..45F}. 

We also observed \cpd\ with  the HARPSpol spectropolarimeter \citep{2011ASPC..437..237S,2011Msngr.143....7P} attached to the ESO 3.6m telescope of La Silla Observatory (Chile) on the 
4th June, 2015. The HARPSpol data cover the spectral range 3780-6910\AA\ with a resolving power of $\approx$\,115\,000. Four exposures of 1800 seconds each were obtained, rotating the quarter-wave retarder plate by 90$^{\circ}$ after each exposure \citep[further details are reported by e.g.][]{2014A&A...564L..10H,Fossati2014}. The final Stokes $I$ spectrum has a signal-to-noise ratio per pixel of approximately 70, calculated at  5000\,\AA.

\section{The strong surface magnetic field of \cpd}
\label{SECT:Mag}

The FORS2 data were reduced and analysed using independent pipelines by the Bonn \citep{2015A&A...582A..45F} and Potsdam \citep{2014MNRAS.440.1779H} teams. The obtained longitudinal magnetic field values (\bzv) are listed in Table~\ref{tab:fors}. Figure~\ref{Fig:FORS_MAGNETIC} shows the output of the Bonn pipeline, where the strong surface longitudinal magnetic field of \cpd\ is  revealed by the  slope in the top-right panel \citep[see Potsdam outcome in][]{2016arXiv161104502S}.

Both pipelines led to a strong magnetic field detection using either the hydrogen lines or the whole spectrum. Nonetheless, there is a difference of approximately 600\,G between the measurements based on the hydrogen lines, whereas the values are consistent within the errors when the whole spectrum is considered. Such discrepancies for strongly magnetic stars are not uncommon \citep[see][]{2014A&A...572A.113L,2015A&A...582A..45F}.

The subsequent HARPSpol spectrum of \cpd\ confirmed the presence of a strong magnetic field. The HARPSpol data were independently reduced and analysed using different  techniques by the Bonn and Potsdam groups, as described in previous BOB works \citep[][]{2014A&A...564L..10H,2016arXiv160102268P}. 

\begin{figure}[]
\resizebox{\hsize}{!}{\includegraphics[angle=0,width=\textwidth]{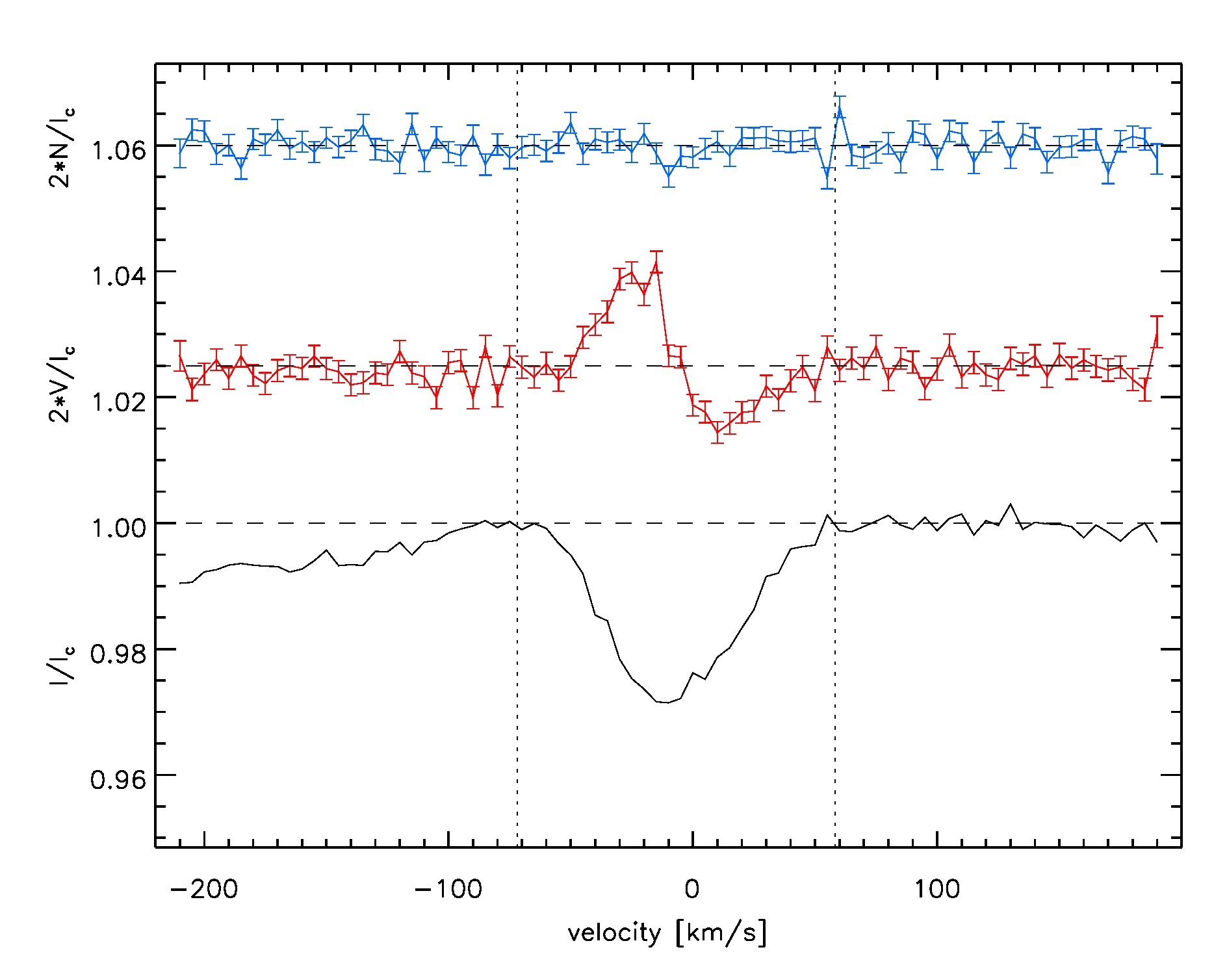}}

\caption{LSD profiles of Stokes $I$ , $V$ and $N$ parameters obtained for \cpd\ by the Bonn group. The vertical dotted lines indicate the velocity range adopted for the determination of the detection probability and \bzv\ value. The Stokes $V$ and $N$ profiles have been shifted upwards by an arbitrary value and expanded by a factor of two.}
\label{Fig:HARPS_MAGNETIC}
\end{figure}

The \bzv\ values, derived from the HARPSpol spectra, were estimated adopting two techniques: the least-squares deconvolution technique \citep[LSD;][]{1997MNRAS.291..658D,2010A&A...524A...5K} and the single value decomposition technique \citep[SVD;][]{2012A&A...548A..95C}. We derived the \bzv\ values  considering 90 metal lines and ignoring He lines because of their large width and the inhomogeneous surface distribution that biases the magnetic field measurements (see Table~\ref{tab:fors}). Each analysis of the Stokes $V$ spectra led to a definite detection of the magnetic field, while, as seen in Fig.~\ref{Fig:HARPS_MAGNETIC}, non-detections were obtained from the diagnostic null profiles \citep[see][for details]{2009PASP..121..993B}. 

The LSD analysis of the rather noisy HARPS observations carried out by the Bonn group resulted in a \bzv\ value of approximately 5.2\,kG. However, the  LSD and SVD applied to the spectra reduced by the Potsdam group led to \bzv\ values of the order of 3-3.5\,kG. On the other hand, if the Potsdam group uses the spectra extracted measurements  of the Bonn group, the  results obtained for SVD and LSD, $5.1\pm0.3$\,kG and $5.4\pm0.3$\,kG, respectively, match the Bonn result well. Since the Potsdam analysis is based on two-dimensional spectra extracted using the ESO HARPS pipeline. It is very likely that the pipeline spectrum extraction is sensitive to the noise level in the observed spectra, which in the case of our HARPS observations is  rather high compared to our previous observations with this instrument, and  for which we have never encountered a similar problem \citep{2014A&A...564L..10H,2015arXiv150703591C,2016arXiv160102268P}. The presence of the remarkably strong magnetic field is also supported by the results of our preliminary reduction of new multi-epoch FORS\,2 observations (Hubrig et al. in preparation) of this star, which indicate that the magnetic field is always stronger than 4 kG.

With a \bzv\  of approximately  5.2\,kG, the magnetic field of \cpd\ is one of the strongest measured so far in stars of the upper main-sequence. Follow-up observations of CPD\,$-62^{\circ}2124$ will constrain its magnetic field strength, its time dependence and geometry and the stellar rotation period.

\begin{table}[h]
\caption[]{Longitudinal magnetic field (\bzv) values of \cpd\ obtained from the FORS2 and HARPSpol spectra.\\[-9mm] }
\label{tab:fors}
\begin{center}
\begin{footnotesize}
\begin{tabular}{ll|lr}
\hline
 \multicolumn{4}{c}{FORS\,2}      \\
\hline
Group & Date/MJD & Lines  & $\langle$B$_z\rangle$\,[G]  \\ 
\hline\hline

  \multirow{2}*{Bonn}  &           &   H lines     & $5220\pm 120$  \\
                       &          \multirow{1}*{17-Mar-2015/    }           &   Whole spectrum    & $4340\pm60$   \\
  \multirow{2}*{Potsdam}  &           57099.21385                       &   H lines    & $4640\pm 130$  \\
                         &                                  &   Whole spectrum    & $4530\pm100$   \\

\hline
 \multicolumn{4}{c}{HARPSpol}      \\

\hline
Group & Date/MJD & Lines &$\langle$B$_z\rangle$ $V$\,[G]   \\
\hline\hline

   \multirow{1}*{Bonn (LSD)}           &          \multirow{1}*{04-Jun-2015/}                         & \multirow{3}*{Metals}                     &    $5200 \pm400$       \\[2pt]

   \multirow{1}*{Potsdam (SVD)}            &        57177.95220                    &     &   $5100\pm300$      \\[2pt]

    \multirow{1}*{Potsdam (LSD)}         &                      &        &     $5400\pm300$     \\[2pt] 

\hline\\[-5mm]

\end{tabular}
\end{footnotesize}
\end{center}
\tablefoot{Column 1 indicates the group that performed the data reduction and analysis. The technique used for the analysis of the HARPSpol spectra is given in brackets. Column 2 lists the observing date. Column 3 gives the considered spectral regions (see Sect.~\ref{SECT:Mag} for details).  } 
\end{table}

\section{Stellar parameters and chemical abundances}
\label{SECT:Para}

 \cpd\ is the second He-strong star for which the BOB consortium discovered the presence of a magnetic field and derived the stellar parameters. \cite{2016arXiv160102268P} presented the magnetic field detection and  parameters for CPD\,$-$57$^{\circ}$\,3509, which is also a He-strong B2IV star, though with a weaker magnetic field (\bzv$\approx$\,1.1\,kG) compared to that of \cpd. 

 The average projected rotational (\ensuremath{{\upsilon}\sin i}) and macroturbulence ($\zeta$) velocities of $35\pm5$\,km\,s$^{-1}$ and $40\pm5$\,km\,s$^{-1}$, respectively, were obtained from the HARPSpol Stokes $I$ spectrum employing the {\sc iacob-broad} code \citep{2014A&A...562A.135S}. The velocities were independently derived using the tools described by \cite{2012A&A...539A.143N}. The two techniques employed approximately 30 and 90 spectral lines, respectively, belonging to nine different chemical elements (see Table 2). The derived $\zeta$  velocity is approximately 1.5 times higher than the average for stars of similar spectral type (Simon-Diaz et al. 2016). Magnetic splitting could  affect the spectral lines, leading to a higher macroturbulence value. Pulsations found in similar  B-type stars could also lead to higher macroturbulence \citep{2014A&A...569A.118A}.

The atmospheric characterisation of \cpd\ was carried out using the hybrid non-LTE approach described by \cite{2007A&A...467..295N} and the same strategy as followed for CPD\,$-$57$^{\circ}$\,3509. A description of the technique, atomic models and limitations is given by \cite{2016arXiv160102268P}, with the difference being that in this analysis we did not consider the hydrogen Balmer lines that are contaminated by circumstellar emission (see Sect.~\ref{Ha}).  The atmospheric parameters were obtained considering the collective ionisation equilibria for elements for which lines are presented in the HARPSpol spectrum: \ion{He}{I/II}, \ion{C}{II/III}, \ion{Si}{II/III/IV}, \ion{O}{I/II} and \ion{S}{II/III}, which results in small uncertainties. Table~\ref{tab:parameters} lists the obtained parameters and chemical abundances, comparing them to the cosmic abundance standard \citep[CAS;][]{2012A&A...539A.143N,2013EAS....63...13P}. Figure~\ref{Fig:HARPS} shows a comparison between the HARPSpol spectrum and the best fitting synthetic model.   A good match is obtained with only small asymmetries/peculiarities present, suggesting that chemical inhomogeneities, such as spots on the surface, are dominating. Exceptions being the Balmer lines, which show an important contribution from the circumstellar material. \cite{1997A&A...324..949Z}, by fitting the Balmer lines, derived a higher effective temperature of 26000\,K and  surface gravity of 4.2\,dex. The authors mention the emission in the Balmer lines as a possible source of uncertainty. In addition, their neglect of metal lines and non-LTE effects contributes to the differences.

  The lines from the investigated chemical species react differently to magnetic broadening (which is unaccounted for in our modelling). While an overall systematic reduction of abundance values can be expected, the effect should be covered by the 1$\sigma$-uncertainties  (see Table 2), as implied by the good match of model and observation and the only slightly higher uncertainties with respect to standard B-stars, in the CAS study for example.

The stellar mass ($M$), radius ($R$), luminosity ($L$), evolutionary age ($\tau$) and fractional main-sequence lifetime ($\tau/\tau_{\rm{MS}}$) were derived from comparisons with stellar evolutionary tracks  by \cite{146E} and \cite{115B} (Table~\ref{tab:parameters}). When considering the   rotating stellar  evolutionary tracks by \cite{115B}, we inferred stellar parameters using the \textsc{bonnsai}\footnote{http://www.astro.uni-bonn.de/stars/bonnsai} tool \citep{Schneider+2014c}. In the relevant mass range, the small differences between the two sets of tracks are attributed mainly to differences in the adopted overshooting parameters \citep[see e.g.][]{2014A&A...570L..13C}. Note that the evolutionary tracks were computed assuming solar abundances, though we expect this to have  a small effect because the He overabundance should be confined to the outermost layers only, implying no influence on the evolution of the star \citep{2016arXiv160102268P}.

\begin{table}[t]
\caption[ ]{Parameters and elemental abundances of \cpd.\\[-9mm]}
\label{tab:parameters}
\begin{center}
\begin{footnotesize}
\begin{tabular}{llll}
\hline
\hline
\multicolumn{3}{l}{Atmospheric parameters:}\\
$T_\mathrm{eff}\,[K]$           & 23650\,$\pm$\,250\\
$\log g$\,[cgs]            & 3.95\,$\pm$\,0.10\\
$y$\,[number fraction]     & 0.35\,$\pm$\,0.02\\
$\xi$\,[km\,s$^{-1}$]                      & 2\,$\pm$\,1\\
$v \sin i$\,[km\,s$^{-1}$]                 & 35\,$\pm$\,5\\
$\zeta$\,[km\,s$^{-1}$]                    & 40\,$\pm$\,5\\[1.3mm]
\hline
\multicolumn{3}{l}{Non-LTE metal abundances:}\\
 & \cpd& CAS \\
$\log$\,(C/H)\,$+$\,12       & 8.29\,$\pm$\,0.06\,(7) & 8.33\,$\pm$\,0.04\\
$\log$\,(N/H)\,$+$\,12       & 7.72\,$\pm$\,0.14\,(17) & 7.79\,$\pm$\,0.04\\
$\log$\,(O/H)\,$+$\,12       & 8.69\,$\pm$\,0.15\,(27) & 8.76\,$\pm$\,0.05\\
$\log$\,(Ne/H)\,$+$\,12      & 8.15\,$\pm$0.09\,(4)  & 8.09\,$\pm$0.05\\
$\log$\,(Mg/H)\,$+$\,12      & 7.37\,(1) & 7.56\,$\pm$\,0.05\\
$\log$\,(Al/H)\,$+$\,12      & 6.19\,$\pm$\,0.06\,(5) & 6.30\,$\pm$\,0.07\\
$\log$\,(Si/H)\,$+$\,12      & 7.77\,$\pm$0.10\,(10) & 7.50\,$\pm$\,0.05 \\
$\log$\,(S/H)\,$+$\,12       & 7.38\,$\pm$\,0.09\,(7) & 7.14\,$\pm$\,0.06\\
$\log$\,(Fe/H)\,$+$\,12      & 7.59\,$\pm$\,0.12\,(7) & 7.52\,$\pm$\,0.03 \\[1.3mm]
\hline
\multicolumn{3}{l}{Fundamental parameters:}\\
                           &  \citet[][]{146E} & \citet[][]{115B} \\%
$M/M_\odot$                & 10.0$\pm$0.4     & $9.8^{+0.6}  _{-0.4}$\\%
$R/R_\odot$                & 5.8$\pm$0.9      &$5.05^{+0.76}  _{-0.66}$ \\%
$\log L/L_\odot$           & 3.98$\pm$0.11    & $3.88^{+0.11}  _{-0.13}$\\%
$\tau$\,[Myr]                     & 16.4$^{+0.9}_{-2.5}$ & $14.54^{+1.04}  _{-2.12}$ \\
$\tau/\tau_\mathrm{MS}$    & 0.65$^{+0.04}_{-0.10}$ & 0.59$^{+0.11}_{-0.12}$\\[.5mm]
\hline\\[-5mm]
\end{tabular}
\tablefoot{The number of lines considered for the determination of the chemical abundances is given in brackets. The cosmic abundance standard  \citep[CAS,][]{2012A&A...539A.143N} in the solar neighbourhood is given for reference, along with data for  Al and S from \citet{2013EAS....63...13P}.\\[-1cm]}
\end{footnotesize}
\end{center}
\end{table}

\begin{figure}[]
\resizebox{\hsize}{!}{\includegraphics[angle=0,width=\textwidth]{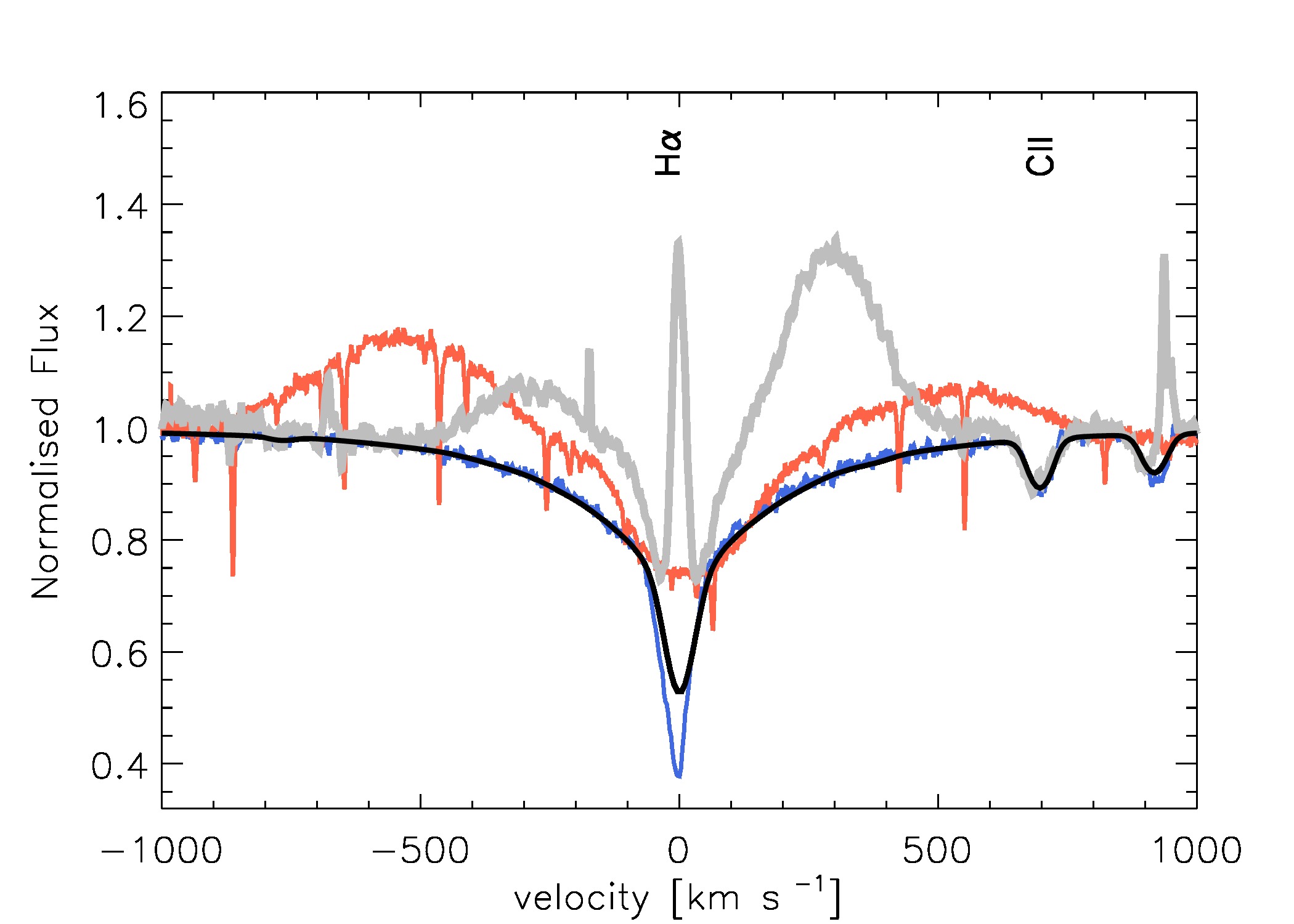}}

\caption{Normalised HARPS  spectrum of \cpd\ (grey solid line) in the H$\alpha$ spectral region. The synthetic photospheric \cpd\ spectrum is also shown (black solid line), together with the HARPSpol spectrum of the He-strong stars CPD\,$-$57$^{\circ}$\,3509 \citep[blue line,][]{2016arXiv160102268P} and that of HD\,37479 \citep[red line, e.g.][]{2012MNRAS.419..959O}. The spectrum of HD\,37479 was obtained within the context of the IACOB project \citep[http://vivaldi.ll.iac.es:8080/iacob/jsp/search.jsp; ][]{2011IAUS..272..310S,2015arXiv150404257S}. The narrow central emission in the \cpd\ spectrum is of nebular origin.}
\label{Fig:Ha}
\end{figure}

\section{Discussion and Conclusion}
\label{Ha}

On the basis of the available information, and assuming a dipolar magnetic field geometry, the lower limit on the dipolar magnetic field strength (B$_{\rm d}$), calculated using the relations of \cite{1967ApJ...150..547P} and assuming a limb darkening coefficient of 0.3 \citep[see][]{2011A&A...529A..75C}, is equal to 18.3\,kG. Note that \cite{2013MNRAS.429..398P} adopted a more conservative limb darkening coefficient of 0.6, under this approximation \cpd's dipolar magnetic field strength is 17.2\,kG. Among the known magnetic massive stars, only NGC\,1624-2 \citep[\bzv=5.35\,kG;][]{2012MNRAS.425.1278W} and the He-strong star HD\,64740 \cite[\bzv$\approx$4.8\,kG;][]{2013MNRAS.429..398P} host a magnetic field with a strength comparable to that of \cpd.

From the measured \ensuremath{{\upsilon}\sin i} value, inferred stellar radius and mass, and assuming an equator-on view, we obtain a maximum rotation period of 7.3\,days and an upper limit on the Keplerian corotation radius of 6.7 stellar radii. Adopting  B$_d$=18.3\,kG, we obtain lower limits on the Alfv{\'e}n radius of approximately 35.2 stellar radii. For these calculations we adopted the stellar parameters obtained from {\sc bonnsai} (including a mass-loss rate of 1.6$\times$10$^{-9}$\,M$_{\odot}$\,yr$^{-1}$) and a terminal velocity of $700$\,km\,s$^{-1}$ \citep{2011MNRAS.416.1456O}. Following the results of \citet{2013MNRAS.429..398P}, \cpd\ should have a centrifugal magnetosphere.

The H$\alpha$ line profile (Fig.~\ref{Fig:Ha}) supports the presence of a disc-like centrifugal magnetosphere. The line presents a central absorption component and line-wing emissions on both sides of the H$\alpha$ line extending up to approximately $\pm300\,$km\,s$^{-1}$. The presence of similar H$\alpha$ emission wings has been  reported for other strongly magnetic He-strong stars, such as  HD\,23478 \citep[\bzv=1.5\,kG;][]{2015A&A...578L...3H} and HD\,37479 \citep[$\sigma$\,Ori\,E, \bzv=2.4\,kG;][]{2012MNRAS.419..959O}, for example. Despite the  similarities between the He-strong stars \cpd\ and CPD\,$-$57$^{\circ}$\,3509, the latter does not show any spectral feature indicative of  circumstellar material (see Fig.~\ref{Fig:Ha}), although on theoretical grounds  it could also host a centrifugal magnetosphere \citep[\bzv=1.1\,kG;][]{2016arXiv160102268P}. \cpd\ therefore provides  a further empirical constrain on the physical conditions (e.g. stellar evolutionary stage, rotation and magnetic field strength) that need to be met for a magnetic star to host and display the signature of a magnetosphere at optical wavelengths \citep{2002ApJ...576..413U,2014arXiv1411.2542S}. The large circumstellar velocities, particularly compared to the low measured \ensuremath{{\upsilon}\sin i} value, may be due to fact that the emission occurs at large stellar radius in conjunction with rigid rotation due to magnetic coupling.  In this scenario, the centrifugal magnetosphere peaks at approximately 8.5 stellar radii.

Based on the B$_d$ value of 18.3\,kG and assuming flux conservation as the only mechanism affecting the magnetic field strength, \cpd\ would have had a dipolar magnetic field strength on the zero-age main-sequence (ZAMS) larger than 34.1\,kG (the evolutionary tracks of \citealt{115B} give a ZAMS radius of 3.7\,R$_{\odot}$). These values could be even larger, if one accounts for the possibility of magnetic field decay \citep{2016arXiv160607599F}. Following the formulation of \citet{2009MNRAS.392.1022U} and \citet{2013MNRAS.429..398P}, and using  the stellar parameters from \textsc{bonnsai}, a gyration constant of 0.04 \citep{2016arXiv160607599F} and the mass-loss rate, terminal velocity and maximum rotation period mentioned earlier in this section, we obtain an upper limit on the spin-down age (the time required to spin down a star from critical rotation to the current rotation rate) of approximately 0.8\,Myr, which is much smaller than the age of the star, suggesting that the magnetic field might have been generated during the main-sequence evolution, possibly following a merger event.

We derived a chemical composition consistent with the CAS \citep{2012A&A...539A.143N,2013EAS....63...13P} and only silicon, sulphur and magnesium (magnesium
abundance based only on \ion{Mg}{ii} $4481\,$\AA){} show small deviations from abundances typical of nearby B stars (off by $\approx0.25$\,dex). The large helium abundance ($y=0.35\pm0.02$\, number fraction) is the only strong peculiarity derived from the abundance analysis. \cpd\ adds to the continuously increasing group of He-strong stars for which a magnetic field has been detected. This further supports the hypothesis \citep{1979ApJ...228..809B} that the rise of a surface He overabundance is closely related to the presence of strong surface magnetic fields.


\begin{acknowledgements}
The authors would like to thank the referee, G. Wade, for his useful comments and suggestions. LF acknowledges financial support from the Alexander von Humboldt Foundation. MFN acknowledges support by the Austrian Science Fund (FWF) in the form of a Meitner Fellowship under project number N-1868-NBL. TM acknowledges financial support from Belspo for contract PRODEX GAIA-DPAC.

\end{acknowledgements}



\bibliographystyle{aa}

\bibliography{AA_2016_29751}

\begin{thebibliography}{55}
\expandafter\ifx\csname natexlab\endcsname\relax\def\natexlab#1{#1}\fi

\bibitem[{{Aerts} {et~al.}(2014){Aerts}, {Sim{\'o}n-D{\'{\i}}az}, {Groot}, \&
  {Degroote}}]{2014A&A...569A.118A}
{Aerts}, C., {Sim{\'o}n-D{\'{\i}}az}, S., {Groot}, P.~J., \& {Degroote}, P.
  2014, \aap, 569, A118

\bibitem[{{Alecian} {et~al.}(2014){Alecian}, {Kochukhov}, {Petit}, {Grunhut},
  {Landstreet}, {Oksala}, {Wade}, {Hussain}, {Neiner}, {Bohlender}, \& {MiMeS
  Collaboration}}]{2014arXiv1404.5508A}
{Alecian}, E., {Kochukhov}, O., {Petit}, V., {et~al.} 2014, \aap, 567, A28

\bibitem[{{Appenzeller} {et~al.}(1998){Appenzeller}, {Fricke}, {F{\"u}rtig},
  {G{\"a}ssler}, {H{\"a}fner}, {Harke}, {Hess}, {Hummel}, {J{\"u}rgens},
  {Kudritzki}, {Mantel}, {Meisl}, {Muschielok}, {Nicklas}, {Rupprecht},
  {Seifert}, {Stahl}, {Szeifert}, \& {Tarantik}}]{1998Msngr..94....1A}
{Appenzeller}, I., {Fricke}, K., {F{\"u}rtig}, W., {et~al.} 1998, The
  Messenger, 94, 1

\bibitem[{{Bagnulo} {et~al.}(2009){Bagnulo}, {Landolfi}, {Landstreet}, {Landi
  Degl'Innocenti}, {Fossati}, \& {Sterzik}}]{2009PASP..121..993B}
{Bagnulo}, S., {Landolfi}, M., {Landstreet}, J.~D., {et~al.} 2009, \pasp, 121,
  993

\bibitem[{{Bagnulo} {et~al.}(2006){Bagnulo}, {Landstreet}, {Mason}, {Andretta},
  {Silaj}, \& {Wade}}]{2006A&A...450..777B}
{Bagnulo}, S., {Landstreet}, J.~D., {Mason}, E., {et~al.} 2006, \aap, 450, 777

\bibitem[{{Bohlender}(1988)}]{1988PhDT........53B}
{Bohlender}, D.~A. 1988, PhD thesis, The university of Western Ontario
  (Canasa).

\bibitem[{{Borra} \& {Landstreet}(1979)}]{1979ApJ...228..809B}
{Borra}, E.~F. \& {Landstreet}, J.~D. 1979, \apj, 228, 809

\bibitem[{{Brott} {et~al.}(2011){Brott}, {de Mink}, {Cantiello}, {Langer}, {de
  Koter}, {Evans}, {Hunter}, {Trundle}, \& {Vink}}]{115B}
{Brott}, I., {de Mink}, S.~E., {Cantiello}, M., {et~al.} 2011, \aap, 530, A115+

\bibitem[{{Carroll} {et~al.}(2012){Carroll}, {Strassmeier}, {Rice}, \&
  {K{\"u}nstler}}]{2012A&A...548A..95C}
{Carroll}, T.~A., {Strassmeier}, K.~G., {Rice}, J.~B., \& {K{\"u}nstler}, A.
  2012, \aap, 548, A95

\bibitem[{{Castro} {et~al.}(2015){Castro}, {Fossati}, {Hubrig},
  {Sim{\'o}n-D{\'{\i}}az}, {Sch{\"o}ller}, {Ilyin}, {Carrol}, {Langer},
  {Morel}, {Schneider}, {Przybilla}, {Herrero}, {de Koter}, {Oskinova},
  {Reisenegger}, {Sana}, \& {BOB Collaboration}}]{2015arXiv150703591C}
{Castro}, N., {Fossati}, L., {Hubrig}, S., {et~al.} 2015, \aap, 581, A81

\bibitem[{{Castro} {et~al.}(2014){Castro}, {Fossati}, {Langer},
  {Sim{\'o}n-D{\'{\i}}az}, {Schneider}, \& {Izzard}}]{2014A&A...570L..13C}
{Castro}, N., {Fossati}, L., {Langer}, N., {et~al.} 2014, \aap, 570, L13

\bibitem[{{Claret} \& {Bloemen}(2011)}]{2011A&A...529A..75C}
{Claret}, A. \& {Bloemen}, S. 2011, \aap, 529, A75

\bibitem[{{Donati} {et~al.}(1997){Donati}, {Semel}, {Carter}, {Rees}, \&
  {Collier Cameron}}]{1997MNRAS.291..658D}
{Donati}, J.-F., {Semel}, M., {Carter}, B.~D., {Rees}, D.~E., \& {Collier
  Cameron}, A. 1997, \mnras, 291, 658

\bibitem[{{Drilling}(1981)}]{1981ApJ...250..701D}
{Drilling}, J.~S. 1981, \apj, 250, 701

\bibitem[{{Ekstr{\"o}m} {et~al.}(2012){Ekstr{\"o}m}, {Georgy}, {Eggenberger},
  {Meynet}, {Mowlavi}, {Wyttenbach}, {Granada}, {Decressin}, {Hirschi},
  {Frischknecht}, {Charbonnel}, \& {Maeder}}]{146E}
{Ekstr{\"o}m}, S., {Georgy}, C., {Eggenberger}, P., {et~al.} 2012, \aap, 537,
  A146

\bibitem[{{Fossati} {et~al.}(2015{\natexlab{a}}){Fossati}, {Castro}, {Morel},
  {Langer}, {Briquet}, {Carroll}, {Hubrig}, {Nieva}, {Oskinova}, {Przybilla},
  {Schneider}, {Sch{\"o}ller}, {Sim{\'o}n-D{\'{\i}}az}, {Ilyin}, {de Koter},
  {Reisenegger}, \& {Sana}}]{Fossati2014}
{Fossati}, L., {Castro}, N., {Morel}, T., {et~al.} 2015{\natexlab{a}}, \aap,
  574, A20

\bibitem[{{Fossati} {et~al.}(2015{\natexlab{b}}){Fossati}, {Castro},
  {Sch{\"o}ller}, {Hubrig}, {Langer}, {Morel}, {Briquet}, {Herrero},
  {Przybilla}, {Sana}, {Schneider}, {de Koter}, \& {BOB
  Collaboration}}]{2015A&A...582A..45F}
{Fossati}, L., {Castro}, N., {Sch{\"o}ller}, M., {et~al.} 2015{\natexlab{b}},
  \aap, 582, A45

\bibitem[{{Fossati} {et~al.}(2016){Fossati}, {Schneider}, {Castro}, {Langer},
  {Simon-Diaz}, {Mueller}, {de Koter}, {Morel}, {Petit}, {Sana}, \&
  {Wade}}]{2016arXiv160607599F}
{Fossati}, L., {Schneider}, F.~R.~N., {Castro}, N., {et~al.} 2016, \aap, 592,
  A84

\bibitem[{{Fossati} {et~al.}(2014){Fossati}, {Zwintz}, {Castro}, {Langer},
  {Lorenz}, {Schneider}, {Kuschnig}, {Matthews}, {Alecian}, {Wade}, {Barnes},
  \& {Thoul}}]{2014A&A...562A.143F}
{Fossati}, L., {Zwintz}, K., {Castro}, N., {et~al.} 2014, \aap, 562, A143

\bibitem[{{Grunhut} {et~al.}(2016){Grunhut}, {Wade}, {Neiner}, {Oksala},
  {Petit}, {Alecian}, {Bohlender}, {Bouret}, {Henrichs}, {Hussain},
  {Kochukhov}, \& {the MiMeS Collaboration}}]{2016arXiv161007895G}
{Grunhut}, J.~H., {Wade}, G.~A., {Neiner}, C., {et~al.} 2016, ArXiv e-prints:
  1610.07895

\bibitem[{{Hubrig} {et~al.}(2014{\natexlab{a}}){Hubrig}, {Fossati}, {Carroll},
  {Castro}, {Gonz{\'a}lez}, {Ilyin}, {Przybilla}, {Sch{\"o}ller}, {Oskinova},
  {Morel}, {Langer}, {Scholz}, {Kharchenko}, \& {Nieva}}]{2014A&A...564L..10H}
{Hubrig}, S., {Fossati}, L., {Carroll}, T.~A., {et~al.} 2014{\natexlab{a}},
  \aap, 564, L10

\bibitem[{{Hubrig} {et~al.}(2015){Hubrig}, {Sch{\"o}ller}, {Fossati}, {Morel},
  {Castro}, {Oskinova}, {Przybilla}, {Eikenberry}, {Nieva}, \&
  {Langer}}]{2015A&A...578L...3H}
{Hubrig}, S., {Sch{\"o}ller}, M., {Fossati}, L., {et~al.} 2015, \aap, 578, L3

\bibitem[{{Hubrig} {et~al.}(2014{\natexlab{b}}){Hubrig}, {Sch{\"o}ller}, \&
  {Kholtygin}}]{2014MNRAS.440.1779H}
{Hubrig}, S., {Sch{\"o}ller}, M., \& {Kholtygin}, A.~F. 2014{\natexlab{b}},
  \mnras, 440, 1779

\bibitem[{{Kochukhov} {et~al.}(2010){Kochukhov}, {Makaganiuk}, \&
  {Piskunov}}]{2010A&A...524A...5K}
{Kochukhov}, O., {Makaganiuk}, V., \& {Piskunov}, N. 2010, \aap, 524, A5

\bibitem[{{Landstreet} {et~al.}(2014){Landstreet}, {Bagnulo}, \&
  {Fossati}}]{2014A&A...572A.113L}
{Landstreet}, J.~D., {Bagnulo}, S., \& {Fossati}, L. 2014, \aap, 572, A113

\bibitem[{{Landstreet} \& {Borra}(1978)}]{1978ApJ...224L...5L}
{Landstreet}, J.~D. \& {Borra}, E.~F. 1978, \apjl, 224, L5

\bibitem[{{Langer}(2012)}]{2012ARA&A..50..107L}
{Langer}, N. 2012, \araa, 50, 107

\bibitem[{{Morel} {et~al.}(2014){Morel}, {Castro}, {Fossati}, {Hubrig},
  {Langer}, {Przybilla}, {Sch{\"o}ller}, {Carroll}, {Ilyin}, {Irrgang},
  {Oskinova}, {Schneider}, {D{\'{\i}}az}, {Briquet}, {Gonz{\'a}lez},
  {Kharchenko}, {Nieva}, {Scholz}, {de Koter}, {Hamann}, {Herrero},
  {Ma{\'{\i}}z Apell{\'a}niz}, {Sana}, {Arlt}, {Barb{\'a}}, {Dufton},
  {Kholtygin}, {Mathys}, {Piskunov}, {Reisenegger}, {Spruit}, \&
  {Yoon}}]{2014arXiv1408.2100M}
{Morel}, T., {Castro}, N., {Fossati}, L., {et~al.} 2014, The Messenger, 157, 27

\bibitem[{{Morel} {et~al.}(2015){Morel}, {Castro}, {Fossati}, {Hubrig},
  {Langer}, {Przybilla}, {Sch{\"o}ller}, {Carroll}, {Ilyin}, {Irrgang},
  {Oskinova}, {Schneider}, {D{\'{\i}}az}, {Briquet}, {Gonz{\'a}lez},
  {Kharchenko}, {Nieva}, {Scholz}, {de Koter}, {Hamann}, {Herrero},
  {Ma{\'{\i}}z Apell{\'a}niz}, {Sana}, {Arlt}, {Barb{\'a}}, {Dufton},
  {Kholtygin}, {Mathys}, {Piskunov}, {Reisenegger}, {Spruit}, \&
  {Yoon}}]{2015IAUS..307..342M}
{Morel}, T., {Castro}, N., {Fossati}, L., {et~al.} 2015, in IAU Symposium, Vol.
  307, 342--347

\bibitem[{{Naz{\'e}} {et~al.}(2010){Naz{\'e}}, {Ud-Doula}, {Spano}, {Rauw}, {De
  Becker}, \& {Walborn}}]{2010A&A...520A..59N}
{Naz{\'e}}, Y., {Ud-Doula}, A., {Spano}, M., {et~al.} 2010, \aap, 520, A59

\bibitem[{{Neiner} {et~al.}(2014){Neiner}, {Tkachenko}, \& {MiMeS
  Collaboration}}]{2014A&A...563L...7N}
{Neiner}, C., {Tkachenko}, A., \& {MiMeS Collaboration}. 2014, \aap, 563, L7

\bibitem[{{Nieva} \& {Przybilla}(2007)}]{2007A&A...467..295N}
{Nieva}, M.~F. \& {Przybilla}, N. 2007, \aap, 467, 295

\bibitem[{{Nieva} \& {Przybilla}(2012)}]{2012A&A...539A.143N}
{Nieva}, M.-F. \& {Przybilla}, N. 2012, \aap, 539, A143

\bibitem[{{Oksala} {et~al.}(2012){Oksala}, {Wade}, {Townsend}, {Owocki},
  {Kochukhov}, {Neiner}, {Alecian}, \& {Grunhut}}]{2012MNRAS.419..959O}
{Oksala}, M.~E., {Wade}, G.~A., {Townsend}, R.~H.~D., {et~al.} 2012, \mnras,
  419, 959

\bibitem[{{Oskinova} {et~al.}(2011){Oskinova}, {Todt}, {Ignace}, {Brown},
  {Cassinelli}, \& {Hamann}}]{2011MNRAS.416.1456O}
{Oskinova}, L.~M., {Todt}, H., {Ignace}, R., {et~al.} 2011, \mnras, 416, 1456

\bibitem[{{Petermann} {et~al.}(2015){Petermann}, {Langer}, {Castro}, \&
  {Fossati}}]{2015A&A...584A..54P}
{Petermann}, I., {Langer}, N., {Castro}, N., \& {Fossati}, L. 2015, \aap, 584,
  A54

\bibitem[{{Petit} {et~al.}(2013){Petit}, {Owocki}, {Wade}, {Cohen},
  {Sundqvist}, {Gagn{\'e}}, {Ma{\'{\i}}z Apell{\'a}niz}, {Oksala}, {Bohlender},
  {Rivinius}, {Henrichs}, {Alecian}, {Townsend}, {ud-Doula}, \& {MiMeS
  Collaboration}}]{2013MNRAS.429..398P}
{Petit}, V., {Owocki}, S.~P., {Wade}, G.~A., {et~al.} 2013, \mnras, 429, 398

\bibitem[{{Piskunov} {et~al.}(2011){Piskunov}, {Snik}, {Dolgopolov},
  {Kochukhov}, {Rodenhuis}, {Valenti}, {Jeffers}, {Makaganiuk}, {Johns-Krull},
  {Stempels}, \& {Keller}}]{2011Msngr.143....7P}
{Piskunov}, N., {Snik}, F., {Dolgopolov}, A., {et~al.} 2011, The Messenger,
  143, 7

\bibitem[{{Preston}(1967)}]{1967ApJ...150..547P}
{Preston}, G.~W. 1967, \apj, 150, 547

\bibitem[{{Przybilla} {et~al.}(2016){Przybilla}, {Fossati}, {Hubrig}, {Nieva},
  {J{\"a}rvinen}, {Castro}, {Sch{\"o}ller}, {Ilyin}, {Butler}, {Schneider},
  {Oskinova}, {Morel}, {Langer}, {de Koter}, \& {BOB
  Collaboration}}]{2016arXiv160102268P}
{Przybilla}, N., {Fossati}, L., {Hubrig}, S., {et~al.} 2016, \aap, 587, A7

\bibitem[{{Przybilla} {et~al.}(2013){Przybilla}, {Nieva}, {Irrgang}, \&
  {Butler}}]{2013EAS....63...13P}
{Przybilla}, N., {Nieva}, M.~F., {Irrgang}, A., \& {Butler}, K. 2013, in EAS
  Publications Series, Vol.~63, 13--23

\bibitem[{{Renson} \& {Manfroid}(2009)}]{2009A&A...498..961R}
{Renson}, P. \& {Manfroid}, J. 2009, \aap, 498, 961

\bibitem[{{Schneider} {et~al.}(2014){Schneider}, {Langer}, {de Koter}, {Brott},
  {Izzard}, \& {Lau}}]{Schneider+2014c}
{Schneider}, F.~R.~N., {Langer}, N., {de Koter}, A., {et~al.} 2014, \aap, 570,
  A66

\bibitem[{{Sch\"oller} {et~al.}(2016){Sch\"oller}, {Hubrig}, {Fossati},
  {Carroll}, {Briquet}, {Oskinova}, {Jarvinen}, {Ilyin}, {Castro}, {Morel},
  {Langer}, {Przybilla}, {Nieva}, {Kholtygin}, {Sana}, {Herrero}, {Barba}, {de
  Koter}, \& {the BOB collaboration}}]{2016arXiv161104502S}
{Sch\"oller}, M., {Hubrig}, S., {Fossati}, L., {et~al.} 2016, ArXiv e-prints:
  1611.04502

\bibitem[{{Shultz} {et~al.}(2014){Shultz}, {Wade}, {Rivinius}, {Townsend}, \&
  {the MiMeS Collaboration}}]{2014arXiv1411.2542S}
{Shultz}, M., {Wade}, G., {Rivinius}, T., {Townsend}, R., \& {the MiMeS
  Collaboration}. 2014, ArXiv e-prints:1411.2542

\bibitem[{{Sim{\'o}n-D{\'{\i}}az} {et~al.}(2011){Sim{\'o}n-D{\'{\i}}az},
  {Castro}, {Garcia}, \& {Herrero}}]{2011IAUS..272..310S}
{Sim{\'o}n-D{\'{\i}}az}, S., {Castro}, N., {Garcia}, M., \& {Herrero}, A. 2011,
  in IAU Symposium, Vol. 272, 310--312

\bibitem[{{Sim{\'o}n-D{\'{\i}}az} \& {Herrero}(2014)}]{2014A&A...562A.135S}
{Sim{\'o}n-D{\'{\i}}az}, S. \& {Herrero}, A. 2014, \aap, 562, A135

\bibitem[{{Sim{\'o}n-D{\'{\i}}az} {et~al.}(2015){Sim{\'o}n-D{\'{\i}}az},
  {Negueruela}, {Ma{\'{\i}}z Apell{\'a}niz}, {Castro}, {Herrero}, {Garcia},
  {P{\'e}rez-Prieto}, {Caon}, {Alacid}, {Camacho}, {Dorda}, {Godart},
  {Gonz{\'a}lez-Fern{\'a}ndez}, {Holgado}, \&
  {R{\"u}bke}}]{2015arXiv150404257S}
{Sim{\'o}n-D{\'{\i}}az}, S., {Negueruela}, I., {Ma{\'{\i}}z Apell{\'a}niz}, J.,
  {et~al.} 2015, in Highlights of Spanish Astrophysics VIII, 576--581

\bibitem[{{Snik} {et~al.}(2011){Snik}, {Kochukhov}, {Piskunov}, {Rodenhuis},
  {Jeffers}, {Keller}, {Dolgopolov}, {Stempels}, {Makaganiuk}, {Valenti}, \&
  {Johns-Krull}}]{2011ASPC..437..237S}
{Snik}, F., {Kochukhov}, O., {Piskunov}, N., {et~al.} 2011, in Astronomical
  Society of the Pacific Conference Series, Vol. 437, Solar Polarization 6, 237

\bibitem[{{ud-Doula} \& {Owocki}(2002)}]{2002ApJ...576..413U}
{ud-Doula}, A. \& {Owocki}, S.~P. 2002, \apj, 576, 413

\bibitem[{{Ud-Doula} {et~al.}(2009){Ud-Doula}, {Owocki}, \&
  {Townsend}}]{2009MNRAS.392.1022U}
{Ud-Doula}, A., {Owocki}, S.~P., \& {Townsend}, R.~H.~D. 2009, \mnras, 392,
  1022

\bibitem[{{Wade} {et~al.}(2014){Wade}, {Grunhut}, {Alecian}, {Neiner},
  {Auri{\`e}re}, {Bohlender}, {David-Uraz}, {Folsom}, {Henrichs}, {Kochukhov},
  {Mathis}, {Owocki}, {Petit}, \& {Petit}}]{2013arXiv1310.3965W}
{Wade}, G.~A., {Grunhut}, J., {Alecian}, E., {et~al.} 2014, in IAU Symposium,
  Vol. 302, 265--269

\bibitem[{{Wade} {et~al.}(2012){Wade}, {Ma{\'{\i}}z Apell{\'a}niz}, {Martins},
  {Petit}, {Grunhut}, {Walborn}, {Barb{\'a}}, {Gagn{\'e}},
  {Garc{\'{\i}}a-Melendo}, {Jose}, {Moffat}, {Naz{\'e}}, {Neiner}, {Pellerin},
  {Penad{\'e}s Ordaz}, {Shultz}, {Sim{\'o}n-D{\'{\i}}az}, \&
  {Sota}}]{2012MNRAS.425.1278W}
{Wade}, G.~A., {Ma{\'{\i}}z Apell{\'a}niz}, J., {Martins}, F., {et~al.} 2012,
  \mnras, 425, 1278

\bibitem[{{Walborn}(1983)}]{1983ApJ...268..195W}
{Walborn}, N.~R. 1983, \apj, 268, 195

\bibitem[{{Zboril} {et~al.}(1997){Zboril}, {North}, {Glagolevskij}, \&
  {Betrix}}]{1997A&A...324..949Z}
{Zboril}, M., {North}, P., {Glagolevskij}, Y.~V., \& {Betrix}, F. 1997, \aap,
  324, 949

\end{thebibliography}


\begin{appendix}
\section{Bonn pipeline output for the FORS2 \cpd\ data}

\begin{figure*}[]
\resizebox{\hsize}{!}{\includegraphics[angle=0,width=\textwidth]{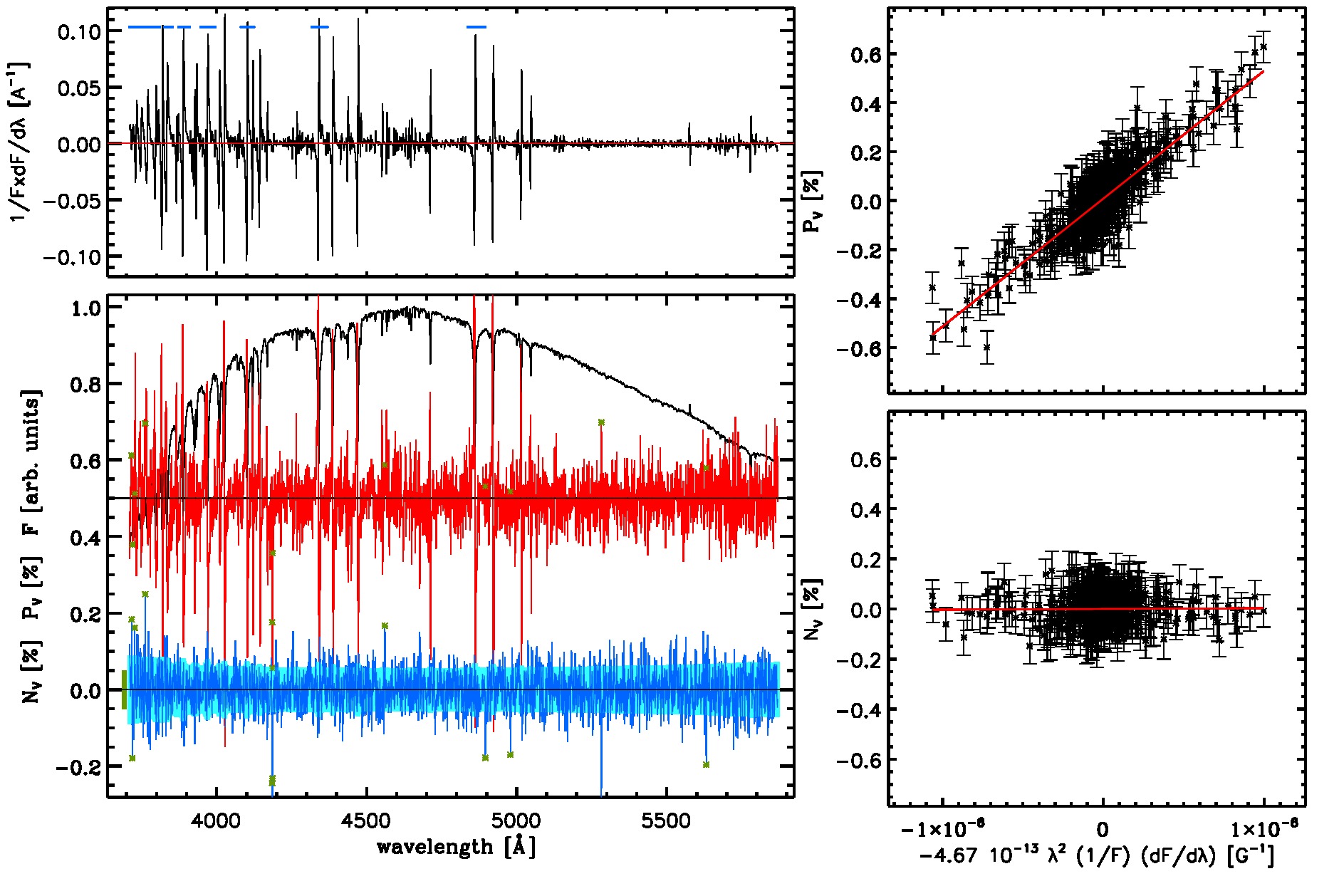}}

\caption{Overview of the results of the analysis of the FORS2 data of \cpd\ using the Bonn pipeline and considering the hydrogen lines. The top-left panel shows the derivative of Stokes $I$. The regions used for the calculation of the magnetic field are marked by a thick blue line close to the top of the panel. Bottom-left panel: the top profile shows Stokes $I$ arbitrarily normalised to the highest value, the middle red profile shows Stokes $V$ (in \%)  shifted upwards for visualisation reasons and the bottom blue profile shows the spectrum of the $N$ parameter (in \%). The green asterisks mark the points that have been removed by the $\sigma$ clipping algorithm. The pale blue strip superimposed upon the $N$ profile shows the uncertainty associated with each spectral point. The thick green bar on the left side of the spectrum of the $N$ parameter shows the standard deviation of the $N$ profile. The top-right panel shows the linear fit used for the determination of the magnetic field using the Stokes $V$ (i.e. \bzv). The red solid line shows the best fit. From the linear fit we obtain \bzv$=5222\pm123$\,G. The bottom-right panel is the same as the bottom-left panel, but for the null profile (i.e. \bzn). From the linear fit, we obtain \bzn$=25\pm95$\,G.}

\label{Fig:FORS_MAGNETIC}
\end{figure*}

\section{\cpd\ HARPSpol stellar atmosphere modelling}

       \begin{figure*}
                \resizebox{\hsize}{!}{\includegraphics[angle=90,width=\textwidth]{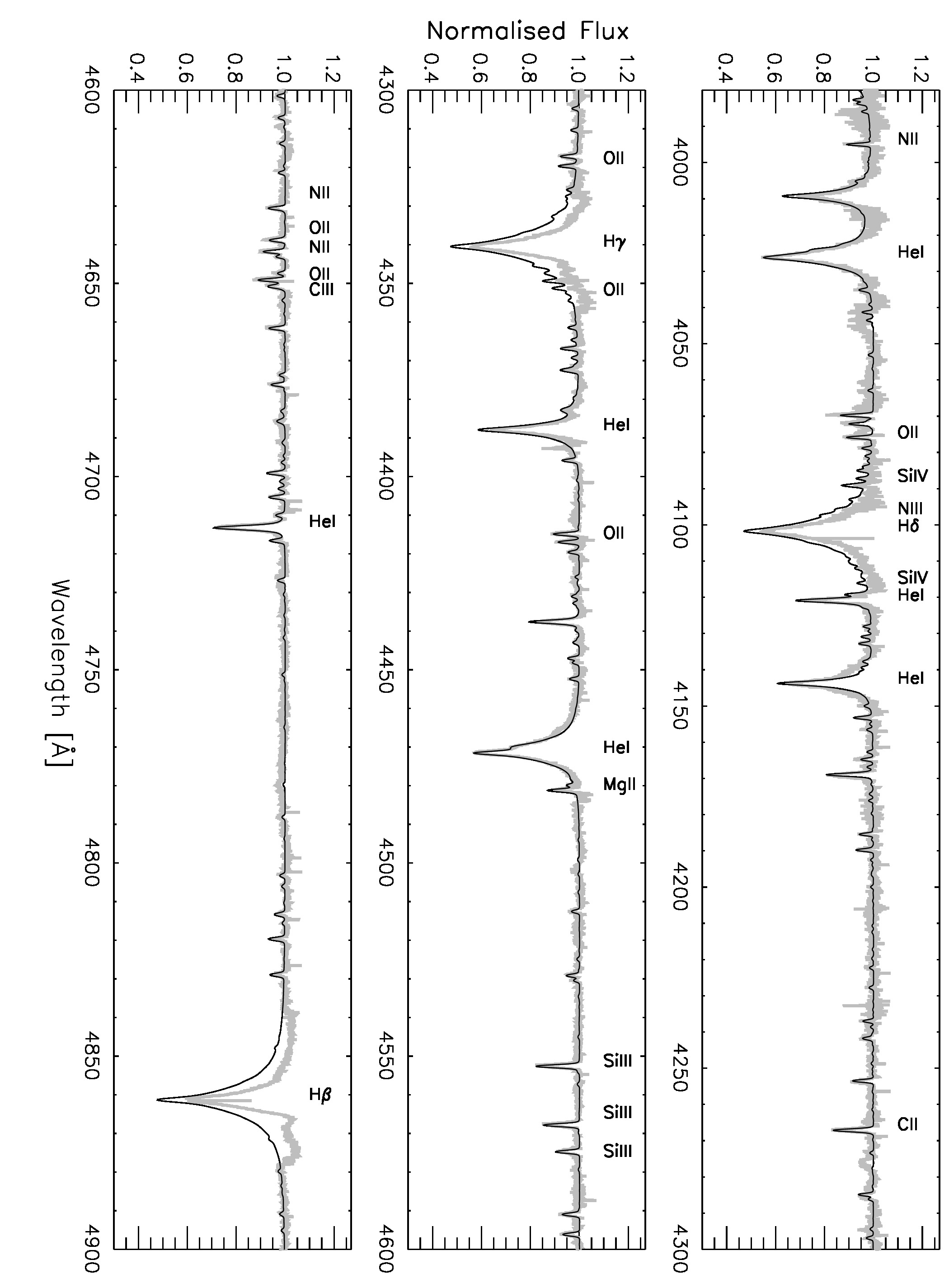}}
                \caption{Normalised HARPSpol optical spectrum (grey) and the best-fitting stellar model (black). Some of the most prominent features are marked.}
                \label{Fig:HARPS}
        \end{figure*}

\end{appendix}

\end{document}